\documentclass[notoc]{JHEP3}
\usepackage{epsfig}
\usepackage{amsmath,bbm}
\usepackage{amssymb}

\newcommand{\ie}{{\it i.e.}}
\newcommand{\eg}{{\it e.g.}}
\newcommand{\cf}{{\it cf.}}

\newcommand{\qu}{{\rm q}}
\newcommand{\qb}{${\rm\bar q}$}
\newcommand{\qbm}{{\rm\bar q}}
\newcommand{\qq}{\qu\qb\ }
\newcommand{\kvec}{\vect{k}}
\newcommand{\pvec}{\vect{p}}

\newcommand{\bs}[1]{\boldsymbol{#1}}
\newcommand{\vect}[1]{\boldsymbol{#1}}

\newcommand{\qv}{\boldsymbol{q}}

\newcommand{\as}{\alpha_s}
\newcommand{\ieps}{i\epsilon}

\newcommand{\M}{{\cal M}}

\newcommand{\lsim}{\buildrel < \over {_\sim}}
\newcommand{\aslash}[1]{ \rlap{/}{#1} }
\newcommand{\Lslash}[1]{ \parbox[b]{1em}{$#1$} \hspace{-0.8em}
                          \parbox[b]{0.8em}{ \raisebox{0.2ex}{$/$}}}

\newcommand{\morder}[1]{{\cal O}\left(#1 \right)}
\newcommand{\eq}[1]{(\ref{#1})}

\newcommand{\halft}{{\textstyle \frac{1}{2}}}
\newcommand{\beq}{\begin{equation}}
\newcommand{\eeq}{\end{equation}}
\newcommand{\nn}{\nonumber}
\newcommand{\beqa}{\begin{eqnarray}}
\newcommand{\eeqa}{\end{eqnarray}}
\newcommand{\be}{\begin{eqnarray}}
\newcommand{\ee}{\end{eqnarray}}
\newcommand{\beqat}{\begin{eqnarray*}}
\newcommand{\eeqat}{\end{eqnarray*}}
\newcommand{\ket}[1]{\vert{#1}\rangle}
\newcommand{\bra}[1]{\langle{#1}\vert}
\newcommand{\ave}[1]{\langle{#1}\rangle}
\newcommand{\com}[2]{\left[#1,#2\right]}

\newcommand{\tr}{\mathrm{Tr}\:}

\newcommand{\um}{\mathbbm{1}}
\newcommand{\ud}{\textrm{d}}
\newcommand{\gf}{\gamma_5}

\newcommand{\Tim}[1]{\mathrm{T}\left\{ {#1} \right\}}

\newcommand{\kb}{{\bar k}}
\newcommand{\ksl}{{\aslash{k}}}
\newcommand{\psl}{{\aslash{p}}}
\newcommand{\qsl}{{\aslash{q}}}
\newcommand{\xsl}{{\aslash{x}}}
\newcommand{\kbsl}{{\aslash{\kb}}}
\newcommand{\dsl}{{\aslash{\partial}}}

\newcommand{\inv}[1]{\frac{1}{#1}}
\newcommand{\csb}{{\raisebox{.5ex}{$\chi$}SB}}

\title{\center{Perturbative gauge theory in a background}}

\author{Dennis D. Dietrich\\The Niels Bohr Institute, Blegdamsvej 17, DK-2100 Copenhagen, Denmark\\
              E-mail: \email{dietrich@nbi.dk}}
\author{Paul Hoyer$^{1,2}$ and Matti J\"arvinen$^{1}$\\
              $^1$Department of Physical Sciences and Helsinki Institute of
              Physics\\
              POB 64, FIN-00014 University of Helsinki, Finland \\
              $^2$NORDITA, Blegdamsvej 17, DK-2100 Copenhagen, Denmark\\
              E-mail: \email{paul.hoyer@helsinki.fi}, \email{Matti.O.Jarvinen@Helsinki.fi}}
\author{St\'ephane Peign\'e\footnote{On leave of absence from LAPTH,
CNRS, UMR 5108, Universit\'e de Savoie, B.P. 110, F-74941
Annecy-le-Vieux Cedex, France.}\\ SUBATECH, UMR 6457, Universit\'e de Nantes \\ Ecole des
Mines de Nantes, IN2P3/CNRS. \\ 4 rue Alfred Kastler, 44307 Nantes
cedex 3, France \\ E-mail: \email{peigne@subatech.in2p3.fr}}
\preprint{\today\\ HIP-2006-30/TH,\ \ NORDITA-2006-18 \\ SUBATECH 2006-03,\ \ LAPTH-1157/06}

\abstract{
Motivated by the gluon condensate in QCD we study the perturbative expansion of a gauge theory in the presence of gauge bosons of vanishing momentum, in the specific case of an abelian theory. The background is characterised by a dimensionful parameter $\Lambda$ affecting only the on-shell prescription of the free (abelian) gluon propagator. When summed to all orders in $g\Lambda$ the modification is equivalent to evaluating standard Green functions in a pure gauge field with an imaginary gauge parameter $\propto \Lambda$. We show how to calculate the corresponding dressed Green functions, which are Poincar\'e and gauge covariant. We evaluate the expressions for the dressed quark and \qq propagators, imposing as boundary condition that they approach the standard perturbative form in the short-distance limit ($|p^2|\to\infty$). The on-shell ($p^2=m^2$) pole of the free quark propagator is removed for any $\Lambda > 0$, and replaced by a discontinuity which vanishes exponentially with $p^2$. The dressing introduces an effective interaction between quarks and antiquarks which is enhanced at low relative 3-momentum. Further study should allow to identify the (bound) eigenstates of propagation and determine whether they define a unitary $S$-matrix.
When the quark mass is zero there is a euclidean propagator solution which breaks chiral symmetry spontaneously.
We study some aspects of the massless pion contribution to axial vector correlators and derive the $\pi$\qq form factor.}


\begin{document}

\section{Introduction}
\label{sec:intro}

The perturbative expansions of field theories such as QED and QCD are defined by the lagrangian once the boundary conditions are specified. It is customary to use the perturbative vacuum $\ket{0}$ as the state around which the expansion is made. This state is empty, as by definition $a_{\kvec}\ket{0}= 0$ for any annihilation operator $a_{\kvec}$. $\ket{in}$ and $\ket{out}$ states with particles are built by applying creation operators on the perturbative vacuum: $a_{\kvec}^\dag\ket{0}$, {\it etc.}

The perturbative expansion works quantitatively for QED to an amazing precision, indicating that the true QED ground state is quite close to the perturbative vacuum. In addition to providing quantitatively reliable results for QED, the perturbative expansion more generally provides analytic and unitary scattering amplitudes -- as expected for $S$-matrix elements in a physical theory.

The perturbative expansion of QCD is qualitatively similar to that of QED. However, we know from experiment that quarks and gluons (unlike electrons and photons) do not appear as asymptotic scattering states. The $S$-matrix of QCD formulated in terms of hadron states should be analytic and unitary. QCD Green functions of quarks and gluons, on the other hand, are not simply related to the $S$-matrix and their analyticity and unitarity properties are largely unknown.

The failure of perturbative QCD to describe the observed hadrons may be caused by the QCD ground state being a quark and gluon condensate. An expansion around the perturbative vacuum would then be inadequate to describe the long distance propagation of quarks and gluons. This need not imply that the coupling $\as$ is large -- in fact, there are plausible arguments \cite{Dokshitzer:1998nz} that the coupling freezes at a moderate value in the long distance regime. The situation may be compared to the propagation of a high energy electron in matter: multiple scattering limits its range even though $\alpha_{em}$ is small.

It is thus interesting to contemplate perturbative expansions around states other than the perturbative vacuum, which might be closer to the ground state of QCD. Such expansions could indicate how an analytic and unitary $S$-matrix is obtained when the asymptotic states are not simply given by the fields of the lagrangian. It is of course desirable that the boundary condition does not break basic symmetries such as Poincar\'e and gauge invariance.

The Feynman $\ieps$ prescription for free quark and gluon propagators is a consequence of expanding around the perturbative vacuum. This may be illustrated by the expression of the free scalar Feynman propagator when the ground state contains a particle of momentum $\kvec$,
\beq\label{kpropdef}
D_k(x-y) \equiv \inv{{\cal N}}\bra{0}a_k\; T[\phi(x) \phi(y)]\; a_k^\dag\ket{0}\ \ ; \hspace{1cm} {\cal N} = \bra{0}a_k\; a_k^\dag\ket{0}
\eeq
which in momentum space is
\beqa\label{kprop}
D_k(p) &=& {\cal{P}}\frac{i}{p^2-m^2} +\pi\delta(p^2-m^2) \nn\\ \\
&+& \frac{(2\pi)^4}{{\cal N}}
\left[ \delta^3 (\pvec-\kvec) \delta(p^0-E_k) +  \delta^3 (\pvec+\kvec) \delta(p^0+E_k) \right] \ \ .\nn
\eeqa
The term in the second line arises from interference between the background and propagating scalars for $p = \pm k$. Only the on-shell propagator is affected, since the background particle is on-shell by definition.

In this paper we study the effect on abelian gauge theory of using an (abelian) gluon propagator of the form
\beq\label{photprop}
D^{\mu\nu}(p) = -g^{\mu\nu}\left[\frac{i}{p^2+\ieps} + \Lambda^2 (2\pi)^4 \delta^4(p) \right] \ \ .
\eeq 
The modification $\propto \Lambda^2$ is analogous to the one in \eq{kprop} for $\kvec=\bs{0}$. Thus the last term in \eq{photprop} signals additional zero-momentum gluons in the $\ket{in}$ and $\ket{out}$ states. We shall not here be concerned with the precise boundary condition implied by this modification of the gluon propagator (see the appendix of \cite{Hoyer:2002ru} for a discussion of this). It turns out, however, that the effect of the modification is formally equivalent to that of a pure gauge field background with an imaginary gauge parameter (see Eq. \eq{field} below). 

In \cite{Hoyer:2004iw,Dietrich:2005ed} we showed that the modification \eq{photprop} can be obtained by assuming a constant background field and averaging over all components of this field with a gaussian weight. It was possible to calculate exactly the dressed quark and gluon propagators in the limit of a large number of colors ($N_c \to\infty$), \ie, for all planar Feynman diagrams. The present study is restricted to the abelian case, and we use a different method to sum the large number of (planar and non-planar) diagrams.

Modifying the (on-shell) gluon propagator only at vanishing momentum ($\pvec = 0$) preserves Poincar\'e invariance and enables us to evaluate exactly the effects of the modification. Thus we get a novel perturbative expansion, with Green functions at each order in $\alpha$ dressed to all orders in $g\Lambda$. We impose as boundary condition that the Green functions reduce to their standard perturbative expressions at short distance ($p^2 \to \pm \infty$). At long distance $|p^2| \lsim (g\Lambda)^2$ they turn out to have a qualitatively different (exponential) behavior.

We view this study as a possible step towards understanding the long-distance properties of QCD. Even though our abelian example is at best a toy model, the explicit results we obtain for the perturbative expansion with a non-trivial boundary condition may be useful for understanding properties of gauge theories in a non-perturbative sector. The vacuum of a perturbative expansion should be such that the essential physics is manifest already at low orders of the expansion. We do not know whether the present approach can fulfill this criterion when applied to QCD. Phenomenological studies \cite{Munczek:1983dx} using modifications of the gluon  propagator similar to \eq{photprop} are, however, encouraging in this regard.

In section 5 we note that for vanishing quark mass there is a quark propagator solution which breaks chiral symmetry (\csb). The symmetry breaking term vanishes exponentially for $p^2 \to -\infty$. Hence the \csb\ propagator gives well-defined results in Wick-rotated (euclidean) loop integrals. Using the methods developed in this paper
we verify the axial Ward identity, study the massless pion contribution to axial vector correlators and derive the $\pi$\qq form factor.

\section{Dressed quark propagator}

In this section we derive the (abelian) quark propagator in the {\it dressed tree} approximation defined in Fig.~1. We sum all (abelian) gluon loops but include only the last term $\propto \Lambda^2$ in each gluon propagator \eq{photprop}. As this part of the propagator carries no momentum the loop integrals are eliminated and the sum can be exactly evaluated.

\EPSFIGURE{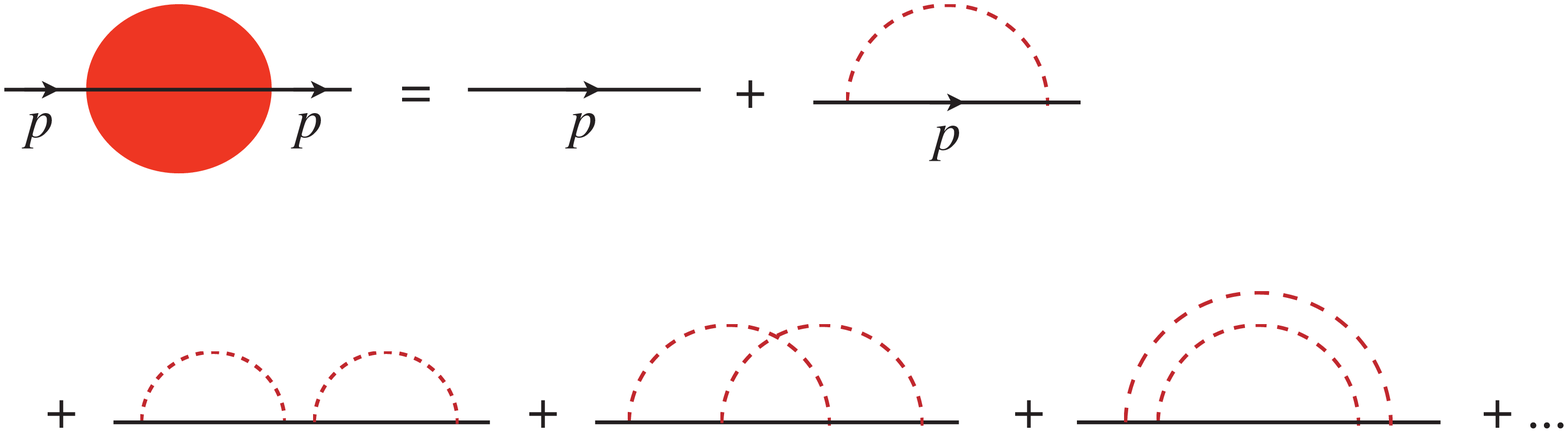,width=.8\columnwidth}{The dressed quark propagator 
to all orders in $g\Lambda$ ({\it lhs.}) is given by the sum of all gluon loop corrections (dashed lines). Only the modification $\propto \Lambda^2$ is included in each gluon propagator \protect{\eq{photprop}}. Self-interactions (via quark loops) of zero-momentum lines are neglected.}

Using standard Feynman rules, the sum shown in Fig.~1 for the dressed tree propagator begins as (an $\ieps$ prescription is implied)
\beq\label{S1}
S(p) = \frac{i}{\psl-m}+\frac{2i(g\Lambda)^2}{(p^2-m^2)^2}\left(\psl-\frac{2m^2}{\psl-m}\right) + \morder{(g\Lambda)^4} \ \ .
\eeq 
Since the parameter $\Lambda$ will always occur in the combination $g\Lambda$ it is convenient to define
\beq\label{mudef}
\mu \equiv g\Lambda \ \ .
\eeq
%
\EPSFIGURE{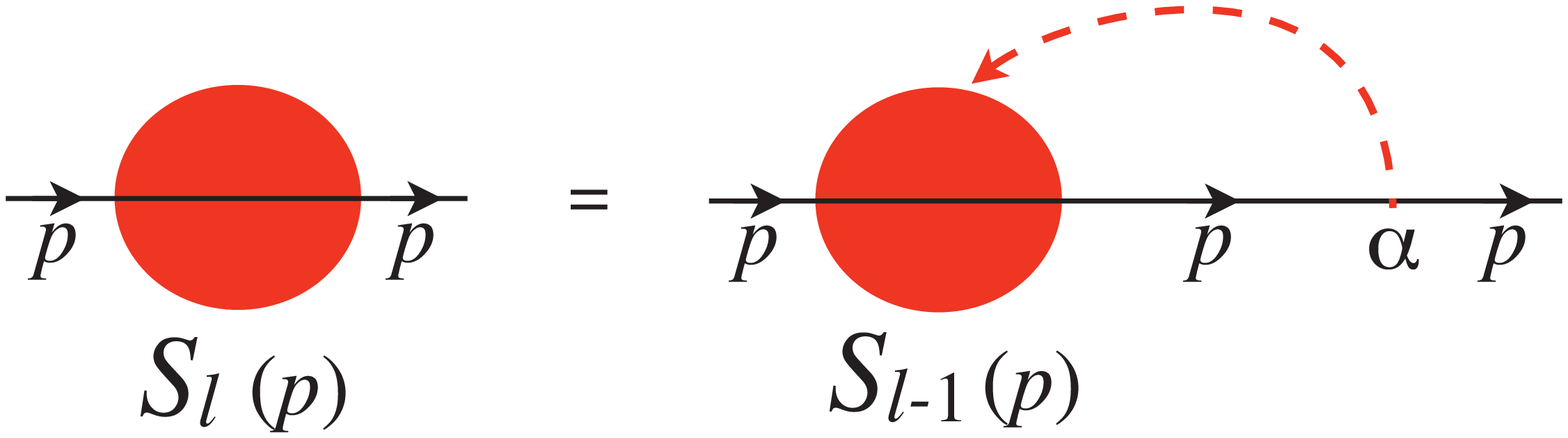,width=.5\columnwidth}{Recurrence relation \eq{recurrency} for the quark propagator dressed with $\ell$ gluon loops. The gluon line starting at the first vertex on the right is to be inserted at all positions between the remaining vertices.}
\noindent
Furthermore, for dimensional reasons the propagator (and any Green function) can be written as a function of the dimensionless ratios $p/\mu$ and $m/\mu$, times the power of $\mu$ corresponding to its dimension. We therefore change our notation as follows:
\beq\label{dimless}
\frac{p}{\mu} \to p\ \ , \hspace{1cm} \frac{m}{\mu}\to m \ \ .
\eeq 
In the following, momenta and masses will be always dimensionless, unless otherwise indicated. With this new notation the dressed tree quark propagator can be expressed as
\beq
\label{Sdef}
S(p) = \sum_{\ell=0}^\infty S_{\ell}(p) \ \ ; \ \  S_0(p) \equiv  \inv{\mu}\, \frac{i}{\psl-m} \ \ ,
\eeq
where $S_{\ell}(p)$ denotes the $\ell$-loop contribution to $S(p)$. 

For $\ell \ge 1$ $S_\ell(p)$ satisfies the recurrence relation shown in Fig.~2,
where the arrow on the (dashed) gluon propagator indicates that it should be inserted at all possible locations in $S_{\ell-1}(p)$. At each location this turns a free propagator $S_0(p)$ into $S_0(p)(-ig\gamma_\alpha)S_0(p)$. According to the identity 
\beq
\label{identity}
-\frac{g}{\mu}\,\frac{\partial}{\partial p^{\alpha}}S_0(p) = S_0(p) \,(-ig \gamma_{\alpha}) \,S_0(p) \ \ ,
\eeq
the same is achieved by differentiating $S_0(p)$. By differentiating $S_{\ell-1}(p)$ all insertions of the gluon propagator are accounted for. Explicitly,
\beq
\label{recurrency}
S_{\ell}(p) = \inv{\mu}\, \frac{i}{\psl-m} \,(-ig \gamma^{\alpha})(-\Lambda^2)\left(-\frac{g}{\mu}\right)\,\frac{\partial}{\partial p^{\alpha}}\, S_{\ell-1}(p) = \inv{\psl-m} \, \dsl_p \, S_{\ell-1}(p) \ \ ,
\eeq
where $\dsl_p \equiv \gamma^{\alpha} \partial/\partial p^{\alpha}$.
From \eq{Sdef} and \eq{recurrency} we find that $S(p)$ satisfies the differential equation
\beq
(\psl -m-\dsl_p)\, S(p) = \frac{i}{\mu} \ \ .
\label{propeq1}
\eeq
This is actually the Dyson-Schwinger equation of Fig.~3, where the dressed \qu\qb$g$ vertex is obtained by differentiating the dressed quark propagator.
\EPSFIGURE{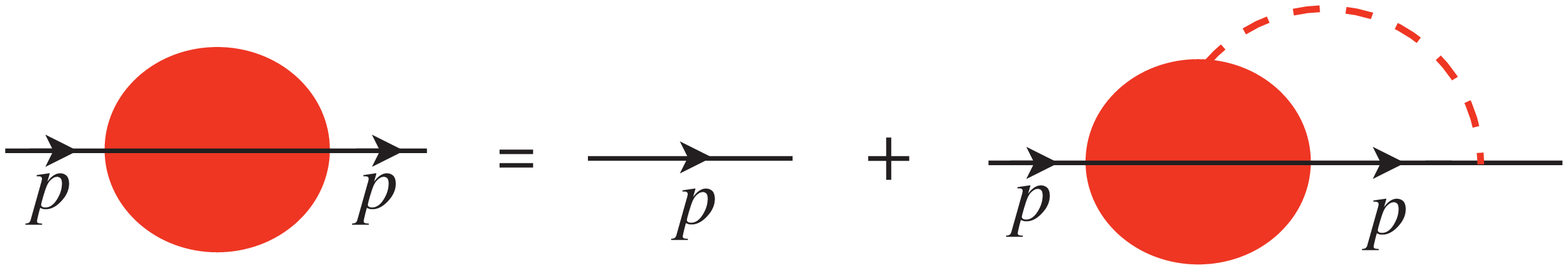,width=.6\columnwidth}{Dyson-Schwinger equation for the dressed quark propagator.}

It is interesting to compare the condition \eq{propeq1} (in coordinate space) with the equation of motion for the propagator in a background field $A^\nu$, which reads\footnote{Recall that we are using dimensionless coordinates, $\mu x \to x$.}
\beq\label{eqmot}
\left[i\dsl_x-m-\frac{g}{\mu}\Lslash{A}(x)\right]S(x)= i\mu^3 \delta^4(x) \ \ .
\eeq
The two previous equations are equivalent for
\beq\label{field}
gA^\nu(x) = i\mu x^\nu = i\mu \partial^\nu(\halft x^2)
\eeq
which formally is a pure gauge with an imaginary gauge parameter. The factor $i$ follows from our modification of the {\it on-shell} gluon propagator in \eq{photprop}, and implies that the field \eq{field} actually influences the physics. The ``gauge transformed'' Green functions differ from the free ones not just by a phase but in absolute size\footnote{This is reminiscent of the original definition by Hermann Weyl in 1918 of gauge symmetry as invariance under a change of scale.}. As a consequence, a general solution of the differential equation \eq{propeq1} (which we shall also call the {\it equation of motion}) for the dressed quark propagator grows exponentially at large $p^2$. We select a physical solution by imposing as boundary condition the perturbative result at $1/p^2 \to 0$, \ie, at large $p^2$ in all directions on the physical sheet.

The equation of motion \eq{propeq1} is transformed into two coupled ordinary differential equations by using Lorentz invariance to express $S(p)$ as
\beq\label{proplor}
S(p) = a(p^2)\aslash{p}+ b(p^2) \ \ .
\eeq 
For Re$\,p^2>0$ the physical solution is given by the integral representation
\beq\label{propsol1}
S(p) = \frac{i}{2\mu}(\psl+m-\dsl_p) \int_0^\infty dt \exp\left[-\frac{t}{2} \left(p^2 - \frac{m^2}{1+t}\right)\right] \ \ .
\eeq
We may check that $S(p)$ satisfies the equation of motion \eq{propeq1} by noting that the operator
\beq\label{opdef}
{\cal O}_p \equiv (\psl-m-\dsl_p)(\psl+m-\dsl_p) = (p-\partial_p)^2-m^2
\eeq
is a scalar in Dirac space and that
\beq\label{oprop}
{\cal O}_p  \exp\left[-\frac{t}{2} \left(p^2 - \frac{m^2}{1+t}\right)\right] = -2\frac{d}{dt} \left\{(1+t)^2 \exp\left[-\frac{t}{2} \left(p^2 - \frac{m^2}{1+t}\right)\right]\right\} \ \ .
\eeq
When inserted into the equation of motion \eq{propeq1} the integrand of the propagator \eq{propsol1} thus turns into an exact derivative. The integral is then trivially evaluated and \eq{propeq1} verified.

It is readily seen from \eq{propsol1} that for ${\rm Re}\,p^2 >m^2$ we recover the standard free quark propagator, either 
in the $\mu \to 0$ limit at fixed (dimensionful) $p^2$ and $m^2$, or in the $p^2 \to \infty$ limit at fixed $\mu^2$ and $m^2$: 
\beq\label{plarge}
S(p) \mathop{\ \longrightarrow \ }_{\mu \to 0}  \frac{i}{\psl -m } \hskip 1cm ; \hskip 1cm  
S(p) \mathop{\ \longrightarrow \ }_{p^2 \to \infty} \frac{i}{\psl}  \ \ .
\eeq
The dressed propagator is regular at $p^2=m^2$ for any $\mu > 0$. The residue of the pole of the free propagator thus vanishes discontinuously with the introduction of a background.

Propagator representations which converge in the upper (Im$\,p^2 > 0$) and lower (Im$\,p^2 < 0$) half planes, respectively, may be found by rotating the integration contour in \eq{propsol1} by $\mp 90^\circ$,
\beq\label{propsol2}
S(p) = \pm\inv{2\mu}(\psl+m-\dsl_p) \int_0^\infty dt \exp\left[\pm \frac{i}{2}t \left(p^2 - \frac{m^2}{1\mp it}\right)\right] \ \ .
\eeq
From these expressions it is apparent that $S(p)$ approaches the free propagator for $|p^2| \to \infty$ in any direction on the $p^2$ plane.

If the integration contour of the representation \eq{propsol1} is rotated by $-180^\circ$
the $t$-integration will range from 0 to $-\infty$ and converge for ${\rm Re}\,p^2<0$. 
The path of integration passes below the essential singularity of the integrand at $t=-1$. This expression is an analytic continuation of $S(p)$ where $p^2$ passes from positive to negative values {\it above} the origin. Similarly, if the contour of \eq{propsol1} is rotated by $+180^\circ$ the contour passes above $t=-1$. The discontinuity of $S(p)$ for $p^2<0$ is given by the difference of the two expressions,
\beq\label{Sdisc}
{\rm Disc}\,S(p^2<0) = -\frac{i}{2\mu}(\psl+m-\dsl_p) \int_{{\cal C}} dt \exp\left[-\frac{t}{2} \left(p^2 - \frac{m^2}{t+1}\right)\right] \ \ ,
\eeq
where $\cal C$ is a closed contour circling the essential singularity at $t=-1$ in the positive (counter-clockwise) direction.

The integral is evaluated by extracting the simple pole contribution at $t=-1$, whose residue turns out to be a Bessel function,
\beq
{\rm Disc}\,S(p^2<0) = -\frac{\pi}{\mu}\,(\psl+m-\dsl_p) \left\{\frac{m}{\sqrt{-p^2}}\,J_1\left(m\sqrt{-p^2}\right) \exp\left[\halft(p^2+m^2)\right]\right\} \ \ .
\eeq
The discontinuity vanishes exponentially for $p^2 \to -\infty$, in agreement with the perturbative propagator having no discontinuity. 
Regarded as an analytic function of $p^2$ the discontinuity is seen to
{\it increase} exponentially for $p^2 \to +\infty$. This 
shows that the propagator is exponentially behaved on sheets other than the physical one.

For $p^2 \to 0$ the dressed propagator \eq{propsol1} behaves as
\beq
 S(p^2)=\frac{ie^{m^2/2}}{\mu}(\psl+m-\dsl_p)
 \left[\inv{p^2}+\frac{m^2}{4}\log p^2+\morder{(p^2)^0} \right] \ \ .
\eeq
Thus the leading singularity $\propto \psl/p^4$ and is exponentially enhanced in $m^2$. We recall that $p$ and $m$ are dimensionless -- thus the above limit corresponds to $p^2 \ll \mu^2$ in terms of dimensionful momentum.

For a massless quark the propagator \eq{propsol1} reduces to
\beq\label{Smassless}
{S(p)\bigg|}_{m=0}= (\psl-\dsl_p) \frac{i}{\mu\,p^2} = 
\frac{i}{\mu\,\psl} \left(1+ \frac{2}{p^2}\right) \ \ .
\eeq
Thus there is only a first order (in $\mu^2$) correction to the massless free propagator. We may verify that for $m=0$ the recurrence relation \eq{recurrency} indeed terminates for $p \neq 0$:
\beq
S_2(p) = \inv{\psl}\,\dsl_p\left(\frac{2i\psl}{\mu\,p^4}\right) = i\frac{4\pi^2}{\mu\,\psl}\,\delta^4(p) \ \ .
\eeq

To conclude this section, it is interesting to note that the scalar integral in the representation \eq{propsol1} of the quark propagator actually corresponds to the quark propagator as evaluated in {\it scalar} abelian gauge theory. We show this in Appendix A, see \eq{scalprop}.
If one recalls the background field interpretation
(see \eq{eqmot} and \eq{field}) this is understandable because for a vanishing 
field strength tensor 
$(i\dsl-\frac{g}{\mu}\Lslash{A}-m)(i\dsl-\frac{g}{\mu}\Lslash{A}+m)=(i\partial-\frac{g}{\mu}A)^2-m^2$, so that the scalar equation of motion follows from the Dirac equation.
The expansion of the scalar integral in \eq{propsol1} in powers of $\mu^2$ is given in \eq{muser} and is asymptotic -- the expansion coefficients grow factorially and the series thus diverges for all values of $\mu$.


\section{Alternative derivation of the dressed quark propagator}


In the previous section we demonstrated that the dressed quark propagator satisfies the differential equation \eq{propeq1}. Making use of its Lorentz structure \eq{proplor} we could solve the equation and find the expression \eq{propsol1} for the propagator. As we shall see in the next section, higher point Green functions satisfy similar equations of motion, but depend on several invariants and have a more involved Lorentz structure. Thus it seems difficult to solve the differential equation even in the case of a double fermion propagator.

Here we present an alternative derivation of the dressed quark propagator, which is easier to apply in the case of higher point Green functions. We make use of the observation that the differential equation \eq{propeq1} is equivalent to the equation of motion of the propagator in a background field \eq{eqmot} which is a pure gauge \eq{field}, albeit with a complex gauge parameter. This allows us to immediately write down a solution of the differential equation. While this solution grows exponentially with $p^2$, we find the physical solution by adding a solution of the homogeneous equation
\beq
(\aslash{p} -m-\dsl_p)\, S_H(p) = 0 \ \ .
\label{homoeq}
\eeq

The equation of motion defined by \eq{eqmot} and \eq{field} in coordinate space,
\beq
(i\dsl_x -m-i\xsl)S(x)=i\mu^3\delta^4(x) \ \ ,
\eeq
is satisfied by
\beq\label{gaugeprop}
S(x) = \exp(\halft x^2)S_0(x) \ \ ,
\eeq
where the free propagator obeys
\beq
(i\dsl_x -m)S_0(x)=i\mu^3\delta^4(x) \ \ .
\eeq
The propagator $S(x)$ grows exponentially with $x^2$ and thus cannot be Fourier transformed to momentum space. In order to have well-defined expressions we temporarily use euclidean metric
and switch back to Minkowski space at the end. Our aim is to find a well-behaved solution of the equation of motion in Minkowski space -- thus the present excursion to euclidean metric may be regarded as a purely technical device.

Making the zeroth component of minkowskian 4-vectors and $\gamma$-matrices imaginary relates their dot products to the euclidean equivalents through substitutions of the form
\beq\label{mintoeu}
p^2 \to -p^2\ ,\hspace{1cm} \psl \to -\psl\ ,\hspace{1cm} \dsl_p \to \dsl_p\ ,\hspace{1cm} \{\gamma^\mu,\gamma^\nu\} = -2\delta^{\mu\nu}\ ,\hspace{1cm} \gf \to \gf \ \ .
\eeq
We shall flag all equations below where the euclidean metric is used by {\it `eucl.'}. Thus the equation of motion
\beq
(i\dsl_x -m+i\xsl)S(x)=i\mu^3\delta^4(x) \hspace{2cm} eucl.
\eeq
is solved by the momentum space propagator
\beqa
S_D(p) &=& \int d^4x e^{-ip\cdot x} \exp(-\halft x^2)\,\frac{i}{\mu} \int\frac{d^4q}{(2\pi)^4} e^{iq\cdot x} \frac{\qsl-m}{q^2+m^2}\nn\\ \nn \\ \label{euprop}
&=& \frac{i}{\mu}\, \int\frac{d^4q}{(2\pi)^2} \exp[-\halft (q-p)^2]\, \frac{\qsl-m}{q^2+m^2} 
 \hspace{2cm} eucl. \\ \nn\\ \label{euprop1}
&=& \frac{i}{\mu}\, (\psl-m+\dsl_p)\, \int\frac{d^4q}{(2\pi)^2} \frac{\exp[-\halft (q-p)^2]}{q^2+m^2} \hspace{1.2cm} eucl.
\eeqa
This dressed euclidean propagator is a gaussian averaged version of the free 
one\footnote{In the notation of our previous work \cite{Hoyer:2004iw,Dietrich:2005ed}, $q^\mu$ would be the background vector potential $\Phi^\mu$.} -- with the smearing scale set by our parameter $\mu$ which defines the dimensionless momenta according to \eq{dimless}. Already from this expression we can see that the propagator will have a bad asymptotic behavior at large $p^2$ in Minkowski metric. Namely, \eq{euprop1} defines $S(p)$ as a function of $p$, which we are free to analytically continue to negative $p^2$. While the integral always converges, the factor $\exp(-\halft p^2)$ in the integrand implies that the propagator grows exponentially for $p^2 \to -\infty\ \ (eucl.)$. The subscript `D' on the propagator \eq{euprop1} indicates that this propagator is distinct from our previous solution \eq{propsol1}.

Introducing the Feynman parametrisation
\beq\label{feynpar}
\inv{q^2+m^2} = \inv{2} \int_0^\infty du\,\exp\left[-\frac{u}{2}(q^2+m^2)\right]
 \hspace{2cm} eucl.
\eeq 
and doing the integral over $q$ gives
\beq
S_D(p) = \frac{i}{2\mu}\, (\psl-m+\dsl_p)\, \int_0^\infty \frac{du}{(1+u)^2} \exp\left[-\frac{u}{2}\left(\frac{p^2}{1+u}+m^2\right)\right] \hspace{1cm} eucl.
\eeq
Finally, we change the integration variable to $t=-u/(1+u)$ and return to Minkowski metric using the translation \eq{mintoeu}. This gives
\beq\label{propsol3}
S_D(p) = \frac{i}{2\mu}(\psl+m-\dsl_p) \int_0^{-1} dt \exp\left[-\frac{t}{2} \left(p^2 - \frac{m^2}{1+t}\right)\right] \ \ .
\eeq
The only difference with our solution \eq{propsol1} is in the upper integration limit. The identity \eq{oprop} ensures that also $S_D(p)$ satisfies the equation of motion \eq{propeq1}, since the integrand vanishes at the endpoint $t=-1$, just as it does at $t=\infty$ for $S(p)$.  However, $S_D(p)$ grows exponentially for $p^2\to\infty$. 

The vanishing of the integrand of \eq{propsol3} at $t=-1$ and $t=\infty$ (for $p^2>0$) implies that
\beq\label{homosol}
S_H(p)=\frac{i}{2\mu}(\psl+m-\dsl_p) \int_{-1}^\infty dt \exp\left[-\frac{t}{2} \left(p^2 - \frac{m^2}{1+t}\right)\right]
\eeq
satisfies the homogenous equation \eq{homoeq}. Our physical solution \eq{propsol1} is given by $S(p) = S_D(p) + S_H(p)$. As we demonstrated in the previous section, $S(p)$ approaches the free propagator as $|p^2| \to \infty$ in all directions on the physical sheet.

\section{Dressed \qq propagator}

We next consider the dressed double propagator of a \qq pair (Fig.~4). Without dressing it is just a direct product of two free propagators,
\beq\label{freeG1}
iG_{0,0}^{\alpha\beta,\rho \sigma}(k,\kb) = \left(\frac{i}{\mu}\,\frac{\ksl+m}{k^2-m^2}\right)^{\alpha \beta}\, \left(\frac{i}{\mu}\,\frac{\kbsl+m}{\kb^2-m^2}\right)^{\rho \sigma} \ \ ,
\eeq
\noindent where the subscripts on the double propagator $G_{n,\bar n}(k,\kb)$ indicate that the number of gluon vertices on the \qu\ line (of momentum $k$) is $n$ and on the \qb\ line (of momentum $-\kb$) is $\bar n$. In order to avoid spelling out the Dirac indices we shall write the direct product in brackets, thus
\beq\label{freeG2}
iG_{0,0} = [S_0(k)]\,[S_0(\kb)] \ \ .
\eeq
\noindent
We wish to evaluate the fully dressed \qq propagator
\beq
G(k,\kb) \equiv \sum_{n=0}^\infty \sum_{\bar n=0}^\infty G_{n,\bar n}(k,\kb) \ \ ,
\eeq
imposing as boundary condition that it approach the free result \eq{freeG1} in the short distance limit. Due to gluon exchanges between the \qu\ and \qb\ lines $G(k,\kb)$ will not factorize into a product of propagators as is the case for $G_{0,0}$ in \eq{freeG2}. Only in the absence of such exchanges we have, \eg,
\beq\label{Gzeron}
iG_{0,\bar n} = [S_0(k)]\,[S_{\bar n}(\kb)] \ \ ,
\eeq
where $S_{\bar n}(\kb)$ is the contribution to $S(\kb)$ which has $\bar n$ vertices on the
\qb\ line.

$G_{n,\bar n}(k,\kb)$ satisfies a recurrence relation which is analogous to that of the single pro\-pa\-ga\-tor in Fig.~2. In diagrams with at least one vertex on the \qu\ line ($n \geq 1$) the gluon propagator from the rightmost vertex must end either to the left on the same line, or on the \qb\ line (Fig.~4). According to \eq{identity} all such insertions are obtained by differentiating the double propagator with two vertices less\footnote{Terms with negative indices are understood to vanish.}:
\beqa\label{Grecur}
iG_{n,\bar n}(k,\kb) &=& \frac{i}{\mu}\,\inv{\ksl-m} (-ig\gamma^\nu)(-\Lambda^2) \left(-\frac{g}{\mu}\right) \left(\frac{\partial}{\partial k^\nu} iG_{n-2,\bar n}(k,\kb) + \frac{\partial}{\partial \kb^\nu} iG_{n-1,\bar n-1}(k,\kb) \right)\nn\\ \\
&=& \frac{i}{\ksl-m}\,\left(\dsl_kG_{n-2,\bar n}(k,\kb) + \dsl_\kb G_{n-1,\bar n-1}(k,\kb)\right)\ \ \ (n \ge 1)  \ \ \ . \nn
\eeqa
We may now sum both sides of \eq{Grecur} over $n$ and $\bar n$, making use of \eq{Gzeron},
\beq
\sum_{n=1}^\infty \sum_{\bar n=0}^\infty G_{n,\bar n}(k,\kb) = G(k,\kb) +i [S_0(k)] \, [S(\kb) ] = \inv{\ksl-m}\,(\dsl_k+\dsl_\kb) G(k,\kb) \ \ .
\eeq 
%
\EPSFIGURE{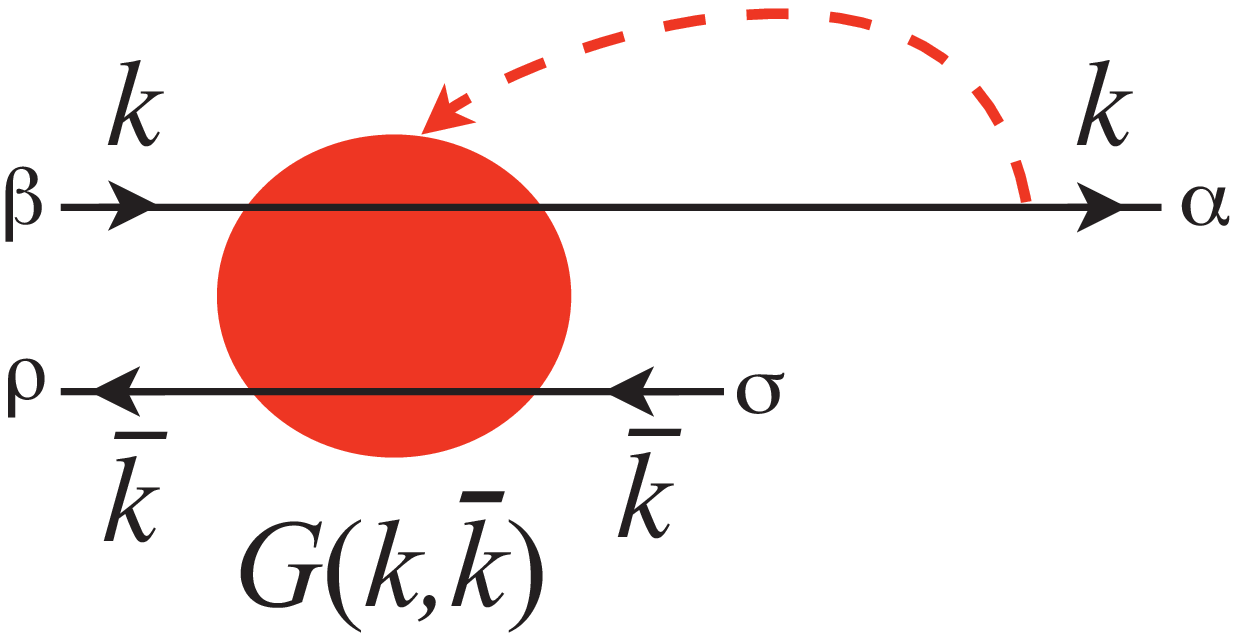,width=.4\columnwidth}{{\it Rhs.} of the recurrence rela\-tion \protect{\eq{Grecur}} for the \qq propagator. Analogous to the {\it rhs.} of Fig.~2.\\ }
%
Here we assumed the momenta $k$ and $\kb$ to be independent. We may instead regard $\kb$ as a dependent variable by keeping the total momentum $p$ of the \qq system,
\beq\label{pdef}
p \equiv k-\bar k
\eeq
fixed when differentiating {\it wrt.} $k$. Then $\dsl_k$ will act also on $\kb = k-p$ and we get the equation of motion (or Dyson-Schwinger equation) for $G$ in the form,
\beq\label{Geom}
(\ksl-m-\dsl_k)_{\alpha\alpha'}G_{\alpha'\beta,\rho\sigma}(k,\kb) = \inv{\mu}\,\delta_{\alpha\beta}\,S_{\rho\sigma}(\kb)
\eeq
where we exhibit the Dirac structure for clarity. 

According to \eq{propeq1}, $\ksl-m-\dsl_k$ is (in an operator sense) the inverse of the dressed propagator $S(k)$. In \eq{Geom} it similarly eliminates the quark line from the \qq propagator $G$. This is non-trivial since (as we shall see below) the dependence of $G(k,\kb)$  on $k$ and $\kb$ cannot be factorized.

Equations analogous to \eq{Geom} where the inverse dressed propagator acts on the other Dirac indices of $G$ can be similarly derived. However, since $G(k,\kb)$ depends on three invariants (which may be taken to be $p^2,k^2$ and $\kb^2$), and many Lorentz structures with four Dirac indices can be built from the two independent momenta $p$ and $k$, it appears difficult to solve the equations of motion directly. We shall instead turn to the alternative method demonstrated in the previous section.

Let us first check that the double propagator $G_D(k,\kb)$ analogous to \eq{euprop} is a solution of the equation of motion \eq{Geom} in euclidean metric (\cf\ \eq{mintoeu}).
\beq \label{GDexpr}
G_D(k,\kb) = \frac{i}{\mu^2} \int \frac{d^4q}{(2\pi)^2} \exp(-\halft q^2)\,\left[\inv{\ksl + \qsl+m}\right]\,\left[\inv{\kbsl + \qsl+m}\right]\hspace{1cm} eucl.
\eeq
As in \eq{freeG2} brackets are used to indicate Dirac structures. The derivative appearing in \eq{Geom} gives
\beqa\label{GDverify}
\dsl_kG_D(k,\kb) &=& \frac{i}{\mu^2} \int \frac{d^4q}{(2\pi)^2} \exp(-\halft q^2)\,\partial_k^\nu \left\{\left[\gamma_\nu\inv{\ksl + \qsl+m}\right]\,\left[\inv{\kbsl + \qsl+m}\right] \right\}\nn \\ \\
&=&\frac{i}{\mu^2} \int \frac{d^4q}{(2\pi)^2} \exp(-\halft q^2)\, \left[\qsl\inv{\ksl + \qsl+m}\right]\,\left[\inv{\kbsl + \qsl+m}\right] \hspace{1cm} eucl. \nn
\eeqa
In the first line the derivative $\partial_k$ (which operates equally on $k$ and $\kb$) can be replaced by $\partial_q$. The result on the second line then follows after a partial integration. Thus \eq{Geom} is indeed satisfied,
\beq
-(\ksl+m+\dsl_k)G_D(k,\kb) = \inv{\mu}\,S_D(\kb) \hspace{2cm} eucl.
\eeq 
with the antiquark propagator on the {\it rhs.} being the euclidean one \eq{euprop}. We anticipate that $G_D$ will not be a solution with good short-distance behavior in Minkowski space, but must be corrected by a solution of the homogeneous equation of motion.

It is helpful to extract the Dirac structure from the integrand as in \eq{euprop1}. Similarly to \eq{GDverify} we note that
\beqa\label{GDirac}
&&\frac{i}{\mu^2} \int \frac{d^4q}{(2\pi)^2} \exp(-\halft q^2)\,\partial_k^\nu \left\{\left[\frac{\gamma_\nu}{(k+q)^2+m^2}\right]\,\left[\inv{\kbsl + \qsl+m}\right] \right\}\nn \\ \\
= &&\frac{i}{\mu^2} \int \frac{d^4q}{(2\pi)^2} \exp(-\halft q^2)\, \left[\frac{\qsl}{(k+q)^2+m^2}\right]\,\left[\inv{\kbsl + \qsl+m}\right] \hspace{1cm} eucl. \nn
\eeqa
It follows that
\beqa
G_D(k,\kb) &=& -\frac{i}{\mu^2}\, [\ksl-m+\dsl_k] \int \frac{d^4q}{(2\pi)^2} \frac{\exp(-\halft q^2)}{(k+q)^2+m^2}\,\left[\inv{\kbsl + \qsl+m}\right] \hspace{4cm} 
\nn \\ \nn \\
 &=& \frac{i}{\mu^2}\, [\ksl-m+\dsl_k]\, [\kbsl-m+\dsl_k] \int \frac{d^4q}{(2\pi)^2} \frac{\exp(-\halft q^2)}{\{(k+q)^2+m^2\}\{(\kb+q)^2+m^2\}} \nn
  \\ \nn \\
 &\equiv&  \frac{i}{\mu^2}\, [\ksl-m+\dsl_k]\, [\kbsl-m+\dsl_k]\, G_{DI}(k^2,\kb^2,p^2)
\hspace{3cm} eucl. \label{gdexpr2}
\eeqa
From the derivation it is clear that $\dsl_k$ in the first bracket acts on the $\kbsl$ in the second bracket as well as on $G_{DI}$.

Introducing Feynman parameters as in \eq{feynpar} and doing the $q$-integrals gives
\beq\label{GIDexpr}
G_{DI}(k^2,\kb^2,p^2) = \inv{4} \int_0^\infty \frac{du\,dv}{(1+u+v)^2}\,\exp\left[-\frac{uk^2 + v\kb^2 + uvp^2}{2(1+u+v)}-\inv{2}(u+v)m^2\right] \ \  eucl.
\eeq
This expression may now be converted back to Minkowski metric using \eq{mintoeu}. At the same time we adjust the integration range as we did for the quark propagator \eq{propsol3}. We note that the integrand in \eq{GIDexpr} vanishes both for $u,v = \infty$ and for $u+v=-1$. Changing the integration range as
\beq
\int_0^\infty du\,dv \to \int_0^{-1} du\,dv\,\theta(1+u+v)
\eeq
will therefore amount to adding  a solution of the homogenous equation of motion. Changing also the signs of $u$ and $v$ we arrive at the Minkowski expression
\beq\label{Gexpr}
G(k,\kb) =  \frac{i}{\mu^2}\, [\ksl+m-\dsl_k]\, [\kbsl+m-\dsl_k]\, G_{I}(k^2,\kb^2,p^2)
\eeq
with
\beq\label{GIexpr}
G_{I}(k^2,\kb^2,p^2) = \inv{4} \int_0^1 du\,dv\, \frac{\theta(1-u-v)}{(1-u-v)^2}\,\exp\left[-\frac{uk^2 + v\kb^2 - uvp^2}{2(1-u-v)}+\inv{2}(u+v)m^2\right] \ \ .
\eeq

The expressions \eq{Gexpr} and \eq{GIexpr} may be regarded as {\it Ans\"atze} for the dressed \qq propagator, motivated by the analysis in euclidean space. We next verify that the equation of motion \eq{Geom} is indeed satisfied. Multiplying $G(k,\kb)$ by $\ksl-m-\dsl_k$ we find the operator ${\cal O}_k$ defined in \eq{opdef}. Since ${\cal O}_k$ commutes with $\kbsl+m-\dsl_k$,
\beq
\left[{\cal O}_k,\kbsl+m-\dsl_k\right] = 0
\eeq
it acts only on $G_{I}(k^2,\kb^2,p^2)$. The identity
\beq\label{opid}
\left({\cal O}_k+2\partial_u \right)\left\{\frac{\theta(1-u-v)}{(1-u-v)^2}\,\exp\left[-\frac{uk^2 + v\kb^2 - uvp^2}{2(1-u-v)}+\inv{2}(u+v)m^2\right]\right\} =0
\eeq
ensures that ${\cal O}_k$ turns the integrand of $G_I$ into a derivative of $u$, so that the $u$-integral is evaluated by substitution. The integral over $v$ assumes the form of the single propagator \eq{propsol1} through the change of variable $t=v/(1-v)$, showing that \eq{Gexpr} is a solution of \eq{Geom}. We note that \eq{GIexpr} is well-defined provided
$uk^2 + v\kb^2 - uvp^2 >0$ when $1-u-v \to 0$. 
This requires $k^2,\, \kb^2 >0$ and $p^2<(\sqrt{k^2} +\sqrt{\kb^2})^2$.

The equation of motion \eq{Geom} ensures the Ward-Takahashi identity for the (non-am\-pu\-ta\-ted) $g$\qq vertex function formed by contracting the $\beta$ and $\rho$ indices in Fig.~4 with $\gamma^\nu$,
\beq\label{Vertex}
\Gamma_{\alpha\sigma}^\nu(k,\kb) \equiv  \frac{i}{\mu^2}\, (\ksl+m-\dsl_k)_{\alpha\beta}\gamma_{\beta\rho}^\nu (\kbsl+m-\dsl_k)_{\rho\sigma}\, G_{I}(k^2,\kb^2,p^2) \ \ .
\eeq
Thus, using \eq{pdef},
\beqa\label{WT}
p_\nu\Gamma^\nu(k,\kb) &=&  \frac{i}{\mu^2}\, (\ksl+m-\dsl_k)\left\{(\ksl-m-\dsl_k)-(\kbsl-m-\dsl_k)\right\} (\kbsl+m-\dsl_k)\, G_{I}(k^2,\kb^2,p^2)  \nn \\ \\
 &=& \frac{1}{\mu}S(\kb) - \frac{1}{\mu}S(k)\nn
\eeqa
follows using \eq{Geom} and the analogous equation of motion with the $\kbsl-m-\dsl_k$ operator. More generally, we expect our dressing to preserve gauge 
symmetry since it corresponds to adding on-shell external (abelian) gluons, whose helicities are contracted using the metric tensor.

Similarly, we may verify the axial Ward identity
\beq\label{awi}
p_\nu \Gamma_5^\nu(k,\kb) = -2im \Gamma_5(k,\kb) + \inv{\mu}S(k)\gamma_5 +\inv{\mu}\gamma_5S(\kb)
\eeq
for the axial vector vertex
\beq\label{axvert}
\Gamma_5^\nu(k,\kb) \equiv \frac{i}{\mu^2}(\ksl+m-\dsl)\,\gamma^\nu\gamma_5\, (\kbsl+m-\dsl)G_I \ \ ,
\eeq
where
\beq\label{psvert}
\Gamma_5(k,\kb) = -\frac{1}{\mu^2}(\ksl+m-\dsl)\, \gf\, (\kbsl+m-\dsl) G_I \ \ .
\eeq

Taking the short distance limit $k^2\to\infty$ and $\kb^2\to\infty$ in \eq{GIexpr}
we have $u \lsim 1/k^2$, $v \lsim 1/\kb^2$ and the integral can be evaluated to give
\beq\label{Gasymp}
iG(k,\kb)\left|_{\buildrel {k^2\to\infty} \over {_{\kb^2\to\infty}}}\right. = \left[ \frac{i}{\mu}\, \frac{\ksl}{k^2} \right]  \ 
\left[ \frac{i}{\mu}\, \frac{\kbsl}{\kb^2} \right] \ \ .
\eeq
To prove the corresponding relation for $k^2,\kb^2\to-\infty$ would require extending the representation analytically. Here we shall only consider the expression for $G(k,\kb)$ in the massless case, $m=0$. Replacing the integration variable $v$ in \eq{GIexpr} by $w=1/(1-u-v)$ the integral over $w$ can be done exactly. A further change of variable to $t=2uk^2/(1-u)$ then gives
\beqa
G_I(k^2,\kb^2,p^2)\left|_{m=0} \right. &=& \int_0^\infty dt \frac{\exp(-t/4)}{t^2+4tk\cdot\kb+4k^2\kb^2} \label{GImassless1}\\ \nn \\
&=& \int_0^\infty dt \frac{\exp(-t/4)}{(t+2k\cdot\kb)^2- \lambda(k^2,\kb^2,p^2)}
\label{GImassless2}
\eeqa
where $\lambda(a,b,c) \equiv a^2+b^2+c^2-2ab-2ac-2bc$. The integral converges for all values of the invariants 
(given a prescription at the singularities of the integrand) and the short distance limit \eq{Gasymp} is seen to follow from \eq{GImassless1} for $k^2,\kb^2$ growing in any direction of their complex planes. 
We recall that $\lambda(k^2,\kb^2,p^2)$ is related to the relative 3-momentum of the \qq pair,
\beq
\label{relmom}
\lambda(k^2,\kb^2,p^2) = 4 \left[( k\cdot p)^2 - k^2 p^2 \right] = 4 (p^0)^2\, \vec{k}^2  \ \ ,
\eeq
where the second equality holds in the rest frame of the \qq pair, with $p=(p^0,\vec{0})$.


\section{Spontaneous chiral symmetry breaking}


The propagator solutions which we discussed so far maintain chiral symmetry for vanishing quark mass, $m=0$. Here 
we consider $m=0$ solutions which spontaneously break the chiral symmetry  of the lagrangian (\csb). 
Because of the presence of dimensionful constants (such as $f_\pi$ and $\varphi_\pi$, see \eq{piavcoupl}
and \eq{pipscoupl}) in the following discussion, for clarity we consider in this section only dimensionful 
momenta (\ie, not scaled by $\mu$ as in \eq{dimless}).

The differential equation \eq{propeq1} for the quark propagator $S(p)=a(p^2)\,\psl +b(p^2)$ is for $m=0$
\beq \label{diffm0}
(\psl - \mu^2 \dsl_p)(a\psl +b)= i \ \ ,
\eeq
implying
\beqa\label{diffm0b}
p^2(a-2 \mu^2 a')-4 \mu^2 a &=& i \nn \\
\psl(b-2 \mu^2 b')=0
\eeqa
where $a' \equiv \ud a(p^2)/\ud p^2$, and similarly for $b(p^2)$.

The solution which approaches the standard perturbative propagator for $|p^2| \to \infty$ is given by \eq{Smassless},
\beq \label{Smassless2}
a(p^2) = \frac{i}{p^2}\left(1+\frac{2 \mu^2}{p^2}\right)\ ; \hspace{1cm} b(p^2)=0
\eeq
The solution of the homogeneous equation \eq{diffm0b} for $b(p^2)$,
\beq \label{csbprop}
S_\chi(p) = b(p^2) = \frac{i}{\mu^2}f_\chi \exp\left(\frac{p^2}{2\mu^2}\right) \ \ ,
\eeq
where $f_\chi \neq 0$ is an arbitrary constant
(with the dimension of mass), 
breaks chiral symmetry and is exponentially suppressed at short distance, $p^2 \to -\infty$. It is not well behaved for Re$\,p^2 \to +\infty$ and was therefore excluded in our previous analysis. 
However, $S_\chi(p)$ is meaningful in euclidean space, \ie, in Wick-rotated loop integrals. Keeping this in mind we next consider some implications of the dressed propagator solution (with $m=0$),
\beq
S(p) = (\psl- \mu^2 \dsl_p)S_I(p) + S_\chi(p)
\eeq
where $S_\chi(p)$ is given by \eq{csbprop} and
\beq
S_I(p) = \frac{i}{p^2}
\eeq
The value of the quark condensate is
\beq\label{qqcond}
\bra{0}\bar\qu \qu\ket{0} = - \int\frac{d^4 p}{(2\pi)^4}\, \tr S(p) = 
4 \frac{f_\chi}{\mu^2} \int \frac{d^4 p}{(2\pi)^4}\, \exp\left(-\frac{p^2}{2\mu^2}\right)=\frac{f_\chi}{\pi^2} \mu^2 \ \ .
\eeq
where in the second equality the integral is Wick rotated to euclidean space, $p^0 \to ip^0$.

It is crucial for an approximation scheme to maintain the axial ward identity (AWI) \eq{awi} \cite{Delbourgo:1979me} in order to correctly describe the physics of \csb\ \cite{Nambu:1961tp}. For gauge theories one such expansion is previously known, based on a systematic truncation of the Dyson-Schwinger equations \cite{Bender:1996bb}. 
Thus we first verify that the AWI is valid in our dressed perturbation theory (at the Born order in which we are working). 

We already noted above that the axial vertex \eq{axvert}, corresponding to the propagator solution with $f_\chi =0$, does satisfy the AWI. In order to construct the axial vertex with $f_\chi \neq 0$ (but $m=0$) we note that in analogy to \eq{euprop}
\beq\label{Sexpr}
S(p) = \frac{1}{\mu^4} \int\frac{d^4q}{(2\pi)^2} \exp\left[-\frac{(q-p)^2}{2\mu^2} \right]\, S_{0}(q) \hspace{1cm} eucl.
\eeq
for the choice of undressed propagator
\beq
S_{0}(q) = i \left[ -\inv{\qsl}+ \mu^2 f_\chi (2\pi)^2 \delta^4(q) \right] \hspace{1.2cm} eucl.
\eeq
where $eucl.$ denotes the euclidean metric \eq{mintoeu}.
The dressed \qq propagator is now obtained by adding the \csb\ contribution to the bare propagators in the expression \eq{GDexpr} of $G_D(k,\kb)$,
\beqa \label{Gchiexpr}
G_{D}(k,\kb)+G_{D\chi}(k,\kb) &=& \nn\\
&& \hspace{-3.5cm} =\frac{i}{\mu^4} \int \frac{d^4q}{(2\pi)^2} e^{-q^2/(2\mu^2)} \left[\frac{-1}{\ksl + \qsl} + \mu^2 f_\chi (2\pi)^2 \delta^4(k+q)\right] \left[\frac{-1}{\kbsl + \qsl} + \mu^2 f_\chi (2\pi)^2 \delta^4(\kb+q)\right]\nn\\
&& \hspace{-3.5cm} = G_D(k,\kb)+ \frac{if_\chi}{\mu^2 p^2}\left\{e^{-\kb^2/(2\mu^2)}\,[\psl]\,[\um]-e^{-k^2/(2\mu^2)}\,[\um]\,[\psl] \right\} \hspace{1cm} eucl.
\eeqa
where $p=k-\kb \neq 0$ is the total momentum of the \qq pair. Returning to Minkowski space according to \eq{mintoeu} we find the additional contribution to the \qq propagator
\beq\label{gchimin}
G_{\chi}(k,\kb) = \frac{if_\chi}{\mu^2 p^2}\left\{e^{\kb^2/(2\mu^2)}\,[\psl]\,[\um]-e^{k^2/(2\mu^2)}\,[\um]\,[\psl] \right\} \ \ ,
\eeq
and to the axial vector vertex \eq{axvert},
\beq\label{axvertchi2}
\Gamma_{5\chi}^\nu(k,\kb) =i\frac{f_\chi}{\mu^2p^2}\gf \left\{p^\nu\left(e^{k^2/(2\mu^2)}+e^{\kb^2/(2\mu^2)}\right) + \halft\com{\gamma^\nu}{\psl}\left(e^{k^2/(2\mu^2)}-e^{\kb^2/(2\mu^2)}\right)\right\} \ \ .
\eeq
The new contributions $S_\chi$ and $\Gamma_{5\chi}^\nu$ to the dressed propagator and vertex are readily seen to satisfy 
the axial Ward identity \eq{awi}.
Similarly to the propagator \eq{csbprop} also the axial vertex increases exponentially for $k^2, \kb^2 \to \infty$, and thus is well behaved only in euclidean space. 

As an application of the \csb\ contribution we evaluate the $\ave{AP}$ correlator between the axial vector $j_5^\nu(x)= \qbm(x)\gamma^\nu\gf\qu(x)$ and pseudoscalar currents,
\beq\label{m5def}
i\M_5^\nu(p) \equiv \int \ud^4x\, e^{ip\cdot (x-y)} \bra{0}\Tim{j_5^\nu(x)\qbm(y)\gf\qu(y)} \ket{0}
\eeq
which vanishes in ordinary and dressed perturbation theory when $m=0$ and $f_\chi=0$. 
The correlator is obtained by contracting the axial vector vertex \eq{axvertchi2} with $\gf$. The loop integral is well-defined after a Wick rotation
\beq\label{m5loop}
\M_5^\nu(p) = \frac{f_\chi}{\mu^2 p^2} \int\frac{\ud^4 k}{(2\pi)^4} \tr\big[e^{-\kb^2/(2\mu^2)}\psl\gf \gamma^\nu\gf 
-e^{-k^2/(2\mu^2)}\gf\psl\gamma^\nu\gf \big]\hspace{1cm} eucl.
\eeq
and we obtain in Minkowski metric
\beq\label{m5result2}
\M_{5}^\nu(p) = -\frac{p^\nu}{p^2+\ieps}\frac{2f_\chi}{\pi^2}\mu^2 \ \ .
\eeq
This expression satisfies the Ward identity of the correlator,
\beq\label{m5wi}
p_\nu\M_5^\nu(p) = - 2\,\bra{0}\qbm(0)\qu(0) \ket{0}
\eeq
given our above result \eq{qqcond} for the quark condensate.

According to \eq{m5result2} $\M_{5}^\nu(p)$ is completely determined by the pion contribution ($p^2=0$ pole) in our dressed Born approximation. The pion is in fact the only single hadron state which couples both to the pseudoscalar and axial vector vertices\footnote{The decay constants $f_{\pi_n}=0$ for radial excitations $\pi_n$ of the pion, see, \eg, \cite{Bhagwat:2006xi}.}. As seen from \eq{m5wi}, the behavior of $\M_{5}^\nu(p)$ at large $p^2$ is given by the quark condensate. This duality between the pion and the quark condensate was noted in \cite{Shifman:1978by}.

It is instructive to evaluate directly the pion contribution to $\M_{5}^\nu(p)$,
\beqa
i\M_{5\pi}^\nu(p) &=& \int\frac{\ud^3\qv}{(2\pi)^32|\qv|}\int \ud^4x e^{ip\cdot (x-y)}\left[\theta(x^0-y^0)\bra{0}j_5^\nu(x)\ket{\pi(q)}\bra{\pi(q)}\qbm(y)\gf\qu(y)\ket{0}\right.\nn\\
&& \hspace{3cm} \left. +\theta(y^0-x^0)\bra{0}\qbm(y)\gf\qu(y)\ket{\pi(q)}\bra{\pi(q)}j_5^\nu(x)\ket{0}\right]
\eeqa
Using the standard parametrization of the pion coupling to the axial vector current,
\beq\label{piavcoupl}
\bra{0}j_5^\nu(x)\ket{\pi(p)} = -ip^\nu f_\pi e^{-ip\cdot x}
\eeq
and denoting the pion coupling to the pseudoscalar current by
\beq\label{pipscoupl}
\bra{0}\qbm(x)\gf\qu(x)\ket{\pi(p)} = i\varphi_\pi  e^{-ip\cdot x}
\eeq
we find
\beq\label{m5result}
\M_{5\pi}^\nu(p) = \frac{p^\nu}{p^2+\ieps} f_\pi \varphi_\pi \ \ .
\eeq

The pion contribution to the (non-amputated) axial vector vertex
\beq\label{axvertex}
i\Gamma_5^\nu(x,x_1,x_2) = \bra{0}\Tim{j_5^\nu(x)\qu(x_1)\qbm(x_2)}\ket{0}
\eeq
is
\beqa\label{axvertpi}
i\Gamma_{5\pi}^\nu(k,\kb) &=& \int\frac{\ud^3\qv}{(2\pi)^3 2|\qv|} (iq^\nu f_\pi)\int\ud^4x_1  \ud^4x_2 e^{ik\cdot(x_1-x)} e^{i\kb\cdot(x-x_2)}\nn\\
&& \times\Big[\theta(x_1^0-x^0) \theta(x_2^0-x^0) e^{iq\cdot x}\bra{0}\Tim{\qu(x_1) \qbm(x_2)}\ket{\pi(q)} \nn\\
&& \hspace{.2cm} -\theta(x^0-x_1^0) \theta(x^0-x_2^0) e^{-iq\cdot x}\bra{\pi(q)}\Tim{\qu(x_1) \qbm(x_2)}\ket{0}\Big]
\eeqa
and involves the non-amputated pion vertex function $V_\pi(q,x_1-x_2)$
\beq\label{Vpidef}
\bra{0}\Tim{\qu_\alpha(x_1) \qbm_\beta(x_2)}\ket{\pi(q)}
 \equiv e^{-iq\cdot(x_1+x_2)/2} (\gf)_{\alpha\beta}\, iV_\pi(q,x_1-x_2)
\eeq
In momentum space we find for the singular (pion pole) 
contributions at $(k-\kb)^2=0$,
\beq\label{g5pires}
\Gamma_{5\pi}^\nu(k,\kb) \equiv \int \ud^4x_1 \ud^4x_2 e^{ik\cdot(x_1-x)-i\kb\cdot(x_2-x)}
\Gamma_{5\pi}^\nu(x,x_1,x_2) = -\frac{\gf f_\pi}{(k-\kb)^2+\ieps} (k-\kb)^\nu\, V_\pi(k,\kb)
\eeq
where
\beq
V_\pi(k,\kb) \equiv \int \ud^4(x_1-x_2) e^{i(k+\kb)\cdot(x_1-x_2)/2}\,V_\pi(q,x_1-x_2)
\eeq
for $k-\kb=q$. Comparing \eq{g5pires} with \eq{axvertchi2} we have
\beq\label{pivert}
V_\pi(k,\kb) = - \frac{i f_\chi}{\mu^2 f_\pi}\left(e^{k^2/(2 \mu^2)}+e^{\kb^2/(2 \mu^2)}\right) \ \ .
\eeq
Thus the $\pi$\qq form factor is exponentially suppressed for $k^2, \kb^2 \to -\infty$.

The pion wave function at the origin \eq{pipscoupl} is
\beqa
\varphi_\pi &=& -i\bra{0}\qbm(0)\gf\qu(0)\ket{\pi(p)} = -4\int\frac{\ud^4 k}{(2\pi)^4} V_\pi(k,\kb) \nn\\ &=& 
-\frac{4f_\chi}{\mu^2 f_\pi}\int\frac{\ud^4 k}{(2\pi)^4} \left(e^{-k^2/(2 \mu^2)}+e^{-\kb^2/(2 \mu^2)}\right)  = -\frac{2f_\chi}{\pi^2 f_\pi}\mu^2 \ \ .
\eeqa
Using this value in \eq{m5result} we see that the single pion intermediate state approximation of the $\ave{AP}$ correlator, $\M_{5\pi}^\nu(p)$, agrees with the dressed perturbative approximation \eq{m5result2}, which was evaluated through a loop integral.

The pion intermediate state contribution to the $\ave{AA}$ correlator\footnote{The time ordering is understood to be the covariant one described in, \eg, Sect. 5-1-7 of \cite{Itzykson:1980rh}. We thank the Referee for bringing this point to our attention.}
\beq\label{aacorr} 
i\,\Pi_5^{\mu\nu}(p) \equiv \int \ud^4x\, e^{ip\cdot (x-y)} \bra{0}\Tim{j_5^\mu(x)j_5^\nu(y)} \ket{0} = \,i\, \Pi_5(p^2)\ \left(p^2 g^{\mu\nu}-p^\mu p^\nu\right)
\eeq
is similarly found to be
\beq\label{aapires}
\Pi_{5\pi}(p^2) =
-\frac{f_\pi^2}{p^2+\ieps}
\eeq
and is transverse due to the choice of covariant time ordering.

The dressed perturbative approximation of the axial vector correlator $\ave{AA}$ \eq{aacorr} is
obtained by contracting 
the Dirac indices of the \qq propagator $G(k,k-p)$ with $\gamma^\mu\gf$ and $\gamma^\nu\gf$ (respectively)
and integrating over the loop momentum $k$. In Minkowski metric and for $m=0$, the dressed \qq propagator
is obtained by adding $G_\chi(k,k-p)$ of \eq{gchimin} to \eq{Gexpr} using \eq{GImassless1},
\beqa
&& G(k,\kb) =  \frac{i}{\mu^4}\, [\ksl-\mu^2\dsl_k]\, [\kbsl-\mu^2\dsl_k]\, G_{I}(k^2,\kb^2,p^2) + G_\chi(k,k-p) 
\label{Gexprzeromass} \\
&& G_{I}(k^2,\kb^2,p^2) =  \int_0^\infty \!\!dt\,\frac{\mu^4 \, e^{-t/4}}{\mu^4 t^2+4\mu^2 t\,k\cdot\kb +4k^2\kb^2} 
\label{GImassless3}
\eeqa
Since the Dirac trace of the chiral odd contribution $G_\chi(k,k-p)$ vanishes it does not contribute to $\ave{AA}$. Using \eq{aacorr} we obtain for $\Pi_5(p^2)$ the dimensionally regularized expression
\beq
\label{pi5def}
i\,\Pi_5(p^2) = \frac{8\lambda^{4-D}}{3p^2\mu^4}\int\frac{d^Dk}{(2\pi)^D}(k-\mu^2\partial_k)\cdot(\kb-\mu^2\partial_k ) 
\, G_{I}(k^2,\kb^2,p^2) 
\eeq
where $\lambda$ is a renormalization scale.
The factor in front of $G_I$ can be expressed in terms of the operator ${\cal O}_k = (k-\mu^2\partial_k)^2$ defined in \eq{opdef},
\beq
(k-\mu^2\partial_k)\cdot(\kb-\mu^2\partial_k )
= (\kb-\mu^2\partial_k ) \cdot (k-\mu^2\partial_k)
= \halft({\cal O}_k + {\cal O}_{\kb}-p^2)
\eeq
Using the identity \eq{opid} in the expression \eq{GIexpr} for $G_I$ we have
\beq
{\cal O}_k G_I(k^2,\kb^2,p^2) = \frac{\mu^4}{\kb^2}
\eeq
This contribution (and the analogous one for ${\cal O}_{\kb}$) vanishes in the dimensionally regulated loop integral \eq{pi5def}, leaving
\beq
i\,\Pi_5(p^2) = -\frac{4}{3\mu^4}\lambda^{4-D}\int\frac{d^Dk}{(2\pi)^D}\, G_{I}(k^2,\kb^2,p^2)
\eeq

The expression \eq{GImassless3} for $G_I$ has $p^2, k^2$ and $\kb^2$-dependent discontinuities which must be specified in the loop integral over $k$. We shall here only be concerned with the leading behavior of $\Pi_5(p^2)$ for $p^2 \to -\infty$, and Wick rotate the integral to euclidean space. Changing the integration variable $t \to 2uk^2/\mu^2$ we then find
\beq
\Pi_5(p^2) = -\frac{2}{3\mu^2}\,\lambda^{4-D}\int\frac{d^Dk}{(2\pi)^D}
\int_0^\infty du\, \frac{\exp\left[ -uk^2/(2\mu^2)\right]}{(uk-\kb)^2}  \hskip 1cm  eucl.
\eeq
At large (euclidean) $p^2$ the loop integral is dominated by large $k^2$, for which the exponential function restricts $u \lsim \mu^2/k^2$. Hence $uk$ can be neglected in the denominator, giving
\beqa
\label{pi5res}
\Pi_5(p^2) &=& -\frac{4}{3}\,\lambda^{4-D}\int\frac{d^Dk}{(2\pi)^D}\,
\inv{k^2 \kb^2} 
+ \morder{\frac{\mu^2}{p^2}}  \nn \\
&=& \inv{6\pi^2}\left[\inv{D-4} + \inv{2} \log\left(\frac{p^2}{\lambda^2}\right)\right] + \morder{\frac{\mu^2}{p^2}} \hskip 1cm  eucl.
\eeqa
with renormalization dependent constants absorbed in the scale $\lambda$.

Although they are suppressed\footnote{This is expected, since the dressed propagators approach the bare ones at high virtuality.} 
by powers of $p^2$ at large $p^2$, there are {\it non-vanishing} effects of the dressing on $\Pi_5(p^2)$, 
as indicated in \eq{pi5res}. 
On the other hand, the {\it unphysical} \qq propagator $G_D$ of \eq{GDexpr} should not affect gauge invariant quantities at any value of $p^2$, since it is obtained through a (complex) gauge transformation analogous to \eq{gaugeprop}.
Its implicitly regularized integral over $k$ can indeed be done exactly by choosing the loop momentum to be $k+q$,
\beqa
\int\frac{d^4 k}{(2\pi)^4} G_D(k,\kb) &=& \frac{i}{\mu^4} \int\frac{d^4 k}{(2\pi)^4} \int\frac{d^4q}{(2\pi)^2} e^{-q^2/(2\mu^2)}
\,\left[\inv{\ksl + \qsl}\right]\,\left[\inv{\ksl - \psl + \qsl}\right] \hspace{.5cm} eucl. \nn \\
&=& \left( \frac{i}{\mu^4} \int \frac{d^4q}{(2\pi)^2} e^{-q^2/(2\mu^2)} \right) \, \int \frac{d^4 k}{(2\pi)^4} 
\,\left[\inv{\ksl}\right]\,\left[\inv{\ksl - \psl}\right] \hspace{1cm} eucl.  \nn \\
&=& i  \, 
\int \frac{d^4 k}{(2\pi)^4} \,\left[\inv{\ksl}\right]\,\left[\inv{\ksl - \psl}\right] \ \ , \hspace{2cm} eucl. 
\eeqa
yielding the bare ($\mu = 0$) result.


\section{Discussion}


We studied some properties of the perturbative expansion of abelian gauge theory
around a vacuum state not being the empty perturbative vacuum but containing (abelian)
gluons with vanishing momenta. We defined the expansion through the modification \eq{photprop} of the on-shell prescription of the gluon propagator. We showed how the effect of this modification can be summed to all orders in the case of the quark and \qq propagators. The generalization of our method to higher order (loop) diagrams as well as to other Green functions appears straightforward since the modification is equivalent to calculating Green functions in the background field \eq{field} which is a pure gauge, albeit with an imaginary gauge parameter. The complex nature of the field reflects the presence of physical gluons in the vacuum state and implies that the dressed Green functions are not equivalent to the standard ones.

Our modified Green functions may be useful for understanding general properties of a confining theory like QCD. They reduce to the standard perturbative result in the short-distance limit, but have essentially different (exponential) behavior at long distance. 
Our results on the short-distance limit are already instructive. The safe limit to take is the one where all invariants are independently large (and spacelike). The expression \eq{GImassless1} for the (Dirac reduced, massless) \qq propagator indeed turns into a product of free propagators when $k^2,\, \kb^2$ and $p^2$ are large. However, if we take $k^2 \to \infty$ keeping $\kb^2$ and $p^2$ fixed the result is the free quark propagator $S_0(k)$ multiplied by a long-distance antiquark propagator which is distinct from \eq{Smassless}, which we derived in the absence of a quark. Apparently the presence of the quark -- even though nearly pointlike -- affects the long-distance propagation of the antiquark.

The dressed quark propagator \eq{propsol1} has no singularities for $p^2 > 0$. Thus the pole of the free propagator at $p^2 = m^2$ is removed no matter how small is the parameter $\Lambda$ in the modification of the gluon propagator \eq{photprop}. This is understandable since the pole arises from an infinite propagation distance, during which even a weak background will matter. Thus the quark is not an asymptotic ($\ket{in}$ or $\ket{out}$) state in the presence of a background. The discontinuity \eq{Sdisc} of its propagator vanishes exponentially for $p^2 \to -\infty$ and is thus only relevant in the ``confinement'' region where the quark cannot be treated as a free particle.

Our assumed background field correlates the quark and antiquark giving their joint propagator \eq{GImassless2} an interesting structure. The kinematic function $\lambda(k^2,\kb^2,p^2)$ vanishes at threshold, $\sqrt{p^2}=\sqrt{k^2} +\sqrt{\kb^2}$, where $k\cdot\kb = -\sqrt{k^2\kb^2} <0$. Then both terms in the denominator of the integrand in \eq{GImassless2} vanish within the integration range. As $p^2$ approaches threshold from below, this singularity leads to a strong potential between the quark and antiquark. Their dynamics becomes effectively non-relativistic since the relative 3-momentum (see \eq{relmom}) is small compared to the ``constituent masses'' given by their virtualities $\sqrt{k^2}$ and $\sqrt{\kb^2}$. 

It will evidently be important for the usefulness of the present study to identify the asymptotic states (``hadrons'') and to verify whether the $S$-matrix formed by them is unitary. In section 5 we took a first step in this direction by noting that there is a quark propagator solution which breaks chiral symmetry spontaneously in the case of vanishing quark mass, and
studied some properties of the massless Goldstone boson.

\acknowledgments
The research of PH and MJ was supported in part by the Academy of Finland through grant 102046. MJ acknowledges a PhD study grant from GRASPANP, the Finnish Graduate School in Particle and Nuclear Physics. The
study of chiral symmetry breaking was mandated by the Referee.

\appendix


\section{Connection to scalar abelian gauge theory}


Above we showed that the Dirac structure of the different solutions for the quark propagator $S$ of \eq{propsol1} and $S_D$ of \eq{propsol3} could be extracted into separate differential operators that operate on scalar functions ($S_I$ and $S_{DI}$, respectively, see \eq{scalprop} below). From \eq{euprop}, \eq{euprop1} we see that in euclidean space $S_{DI}$ is obtained from the free {\it scalar} propagator by Gaussian smearing whereas $S_D$ is the smeared {\it quark} propagator. This suggests that $S_I$ and $S_{DI}$ are solutions for the dressed scalar propagator in scalar abelian gauge theory. In this appendix we show that this is the case by checking that $S_I$ (and $S_{DI}$) satisfy the corresponding equation of motion.

The equation of motion for the scalar propagator $S_I$ can be derived similarly to Eq.~\eq{propeq1}.  The Dyson-Schwinger equation 
for the dressed scalar propagator is shown in Fig.~5. 
The operator $-g\partial_p^\alpha/\mu$ of Eq.~\eq{identity} may be used to write the equation in an analytic form also in the scalar case. As
\beqa \label{doid}
 -\frac{g}{\mu}\partial_p^\alpha\inv{\mu^2}\frac{i}{p^2-m^2} &=& 
\inv{\mu^2}\frac{i}{p^2-m^2}\left(-2ig\mu p^\alpha\right)\inv{\mu^2}\frac{i}{p^2-m^2} \nn\\
\\
 -\frac{g}{\mu}\partial_p^\alpha \left(-2ig\mu p^\beta\right) &=& 2ig^2 g^{\alpha\beta}\nn
\eeqa
all gluon insertions are generated: the first line of \eq{doid} represents an insertion to a scalar propagator and the second line gives an insertion to a one-gluon vertex. The analytic form of the Dyson-Schwinger equation is thus
\beqa \label{DSeq}
 S_I(p) &=& \inv{\mu^2}\frac{i}{p^2-m^2} + \inv{\mu^2}\frac{i}{p^2-m^2}(-2ig\mu p_\alpha)\left(-\Lambda^2\right)\left(-\frac{g}{\mu}\partial_p^\alpha\right)S_I(p) \nn\\
&& + \inv{2}\inv{\mu^2}\frac{i}{p^2-m^2}(2ig^2g_{\alpha\beta})\left(-\Lambda^2\right)^2\left(-\frac{g}{\mu}\partial_p^\alpha\right)\left(-\frac{g}{\mu}\partial_p^\beta\right)S_I(p)\nn\\
&&+\inv{2}\inv{\mu^2}\frac{i}{p^2-m^2}(2ig^2g_{\alpha\beta})\left(-\Lambda^2\right)g^{\alpha\beta}S_I(p)
\eeqa

%
\EPSFIGURE[h]{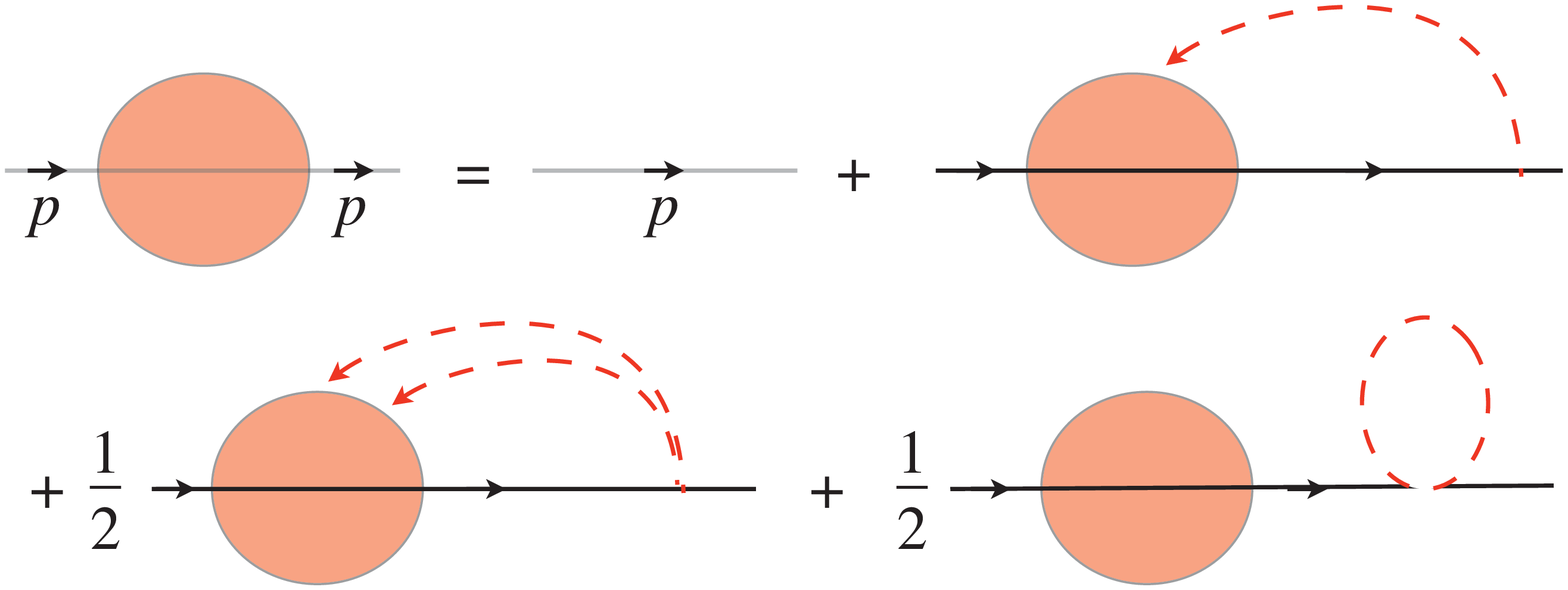,width=.5\columnwidth}{The Dyson-Schwinger equation \protect{\eq{DSeq}} for the dressed scalar propagator.}
%

The extra factors of $1/2$ appearing in the last two terms of \eq{DSeq} are either symmetry factors or required to avoid double coun\-ting of diagrams caused by the second order differential operator. \eq{DSeq} simplifies to
\beq \label{seom}
 \left[(p-\partial_p)^2-m^2\right]S_I(p) = {\cal O}_p\ S_I(p) = \frac{i}{\mu^2}
\eeq
Thus using \eq{oprop} it is easy to check that the scalar part in the quark propagator \eq{propsol1}, 
\beqa \label{scalprop}
S_I(p) &=& \frac{i}{2\mu^2} \int_0^\infty dt \exp\left[-\frac{t}{2} \left(p^2 - \frac{m^2}{1+t}\right)\right] \ \ ,\nn\\&& 
\eeqa
indeed satisfies \eq{seom}. The discussion of section 2 shows that it has the correct short distance behavior. Similarly, one may check that $G_I$ of \eq{GIexpr} is the dressed two-scalar propagator.

It is possible to derive an explicit expression for the perturbation series of Fig.~5.
Restoring the $\mu$ dependence, \eq{scalprop} reads
\beqa
 S_I(p) &=& \frac{i}{2} \int_0^\infty dt \exp\left[-\frac{t}{2} \left(p^2 - \frac{m^2}{1+t\mu^2}\right)\right] \nn\\
    &=& \frac{i}{2} \int_0^\infty dt \exp\left[-\frac{t}{2} \left(p^2 - m^2\right)\right]\exp\left[-\frac{\mu^2t^2m^2}{2\left(1+t\mu^2\right)}\right]
\eeqa
where we scaled $t \to t\mu^2$. Expanding the integrand at $\mu^2=0$ and doing the integration we find
\beqa \label{muser}
 S_I(p) &=& \frac{i}{p^2-m^2} -\frac{4 i m^2 \mu^2}{(p^2-m^2)^3} + \frac{24 i m^2(m^2+p^2)\mu^4}{(p^2-m^2)^5} - \cdots \nn\\ \\
 &=& \frac{i}{p^2-m^2}\left\{1+\frac{m^2}{p^2}\sum_{k=1}^\infty \frac{(-2)^k\, k!}{(1-m^2/p^2)^{2k}} \left(\frac{\mu^2}{p^2}\right)^k\, \sum_{\ell=0}^{k-1} {k+1 \choose \ell+1} {k-1 \choose \ell} \left(\frac{m^2}{p^2}\right)^\ell \right\} \nn
\eeqa
The coefficients in \eq{muser} behave as $\sim k!$, whence the series is asymptotic (unless $m=0$). This is expected, as $S_I$ has a branch point at $\mu^2 = 0$ (see \eq{Sdisc}).

\providecommand{\href}[2]{#2}\begingroup\raggedright
\endgroup

\end{document}